\documentclass[10pt,conference]{IEEEtran}

\usepackage[T1]{fontenc}
\usepackage{microtype}
\usepackage{amsmath,amssymb}
\usepackage{booktabs}
\usepackage{tabularx}
\usepackage{graphicx}
\graphicspath{{figures/}}
\usepackage{xcolor}
\usepackage{tikz}
\usetikzlibrary{positioning,arrows.meta,calc}
\usepackage{algorithm}
\usepackage{algpseudocode}
\usepackage{listings}
\usepackage{cite}
\usepackage{hyperref}
\hypersetup{hidelinks}

\newif\ifqceversion
\qceversiontrue

\title{QuWARP: A Workload-Aware Reuse Planner for Simulating Quantum Circuits}
\author{
  \IEEEauthorblockN{Tim Littau}
  \IEEEauthorblockA{Delft University of Technology\\
  Delft, The Netherlands\\
  t.m.littau@tudelft.nl}
  \and
  \IEEEauthorblockN{Rihan Hai}
  \IEEEauthorblockA{Delft University of Technology\\
  Delft, The Netherlands\\
  r.hai@tudelft.nl}
}

\begin{document}
\maketitle

\begin{abstract}
Quantum circuit simulation often appears as repeated-run workloads rather than isolated circuits: variational quantum eigensolver (VQE) sweeps, noisy multishot studies, and quantum error correction (QEC) cycles revisit closely related structure across many runs. Existing simulators optimise individual executions well, but they largely ignore cross-task shared state and therefore repeat work that could be reused safely. We propose QuWARP, a planner-based workload optimiser for a bounded state of the art simulator execution surface: it performs workload-level planning over related tasks, identifies shared prefixes, and chooses when to materialize exact typed boundary artifacts for later reuse across the evaluated statevector mode, and stabilizer-hybrid mode. Its planner treats continuation legality as a narrow correctness guardrail, abstains when reuse is unprofitable, and keeps each reuse, abstention, or refusal decision auditable through EXPLAIN-style traces, meaning inspectable planner reports with provenance and realized-cost summaries. Across real world quantum application workloads QuWARP delivers $2.95\times$--$32.84\times$ speedups over this work's main direct per-task Qrack denominator on reuse-positive workloads. These results show workload-level reuse planning improves repeated-run simulation while keeping unsupported handoffs auditable and out of the execution path.
\end{abstract}


\section{Introduction}

 This work tackles the problem of avoiding redundant computation across related quantum-simulation runs. It matters because the contribution is not another faster simulator kernel; it is a workload planner that decides when related runs should share exact reusable state and when they should not.

Repeated-run quantum circuit simulation often arrives as a workload of related circuits rather than as one isolated run. Variational quantum eigensolver (VQE) parameter sweeps~\cite{peruzzo2014vqe}, noisy multishot studies, and quantum error correction (QEC) syndrome cycles~\cite{fowler2012surfacecodes} all revisit closely related prefixes across many executions.

In these workloads, much of the wall-clock time is spent re-executing the same shared prefixes even when only late parameters, measurements, or correction steps change. Existing simulators optimise individual runs well, but they usually treat neighbouring runs too independently and therefore redo work that could be reused safely.

\begin{figure}[t]
  \centering
  \includegraphics[width=\columnwidth]{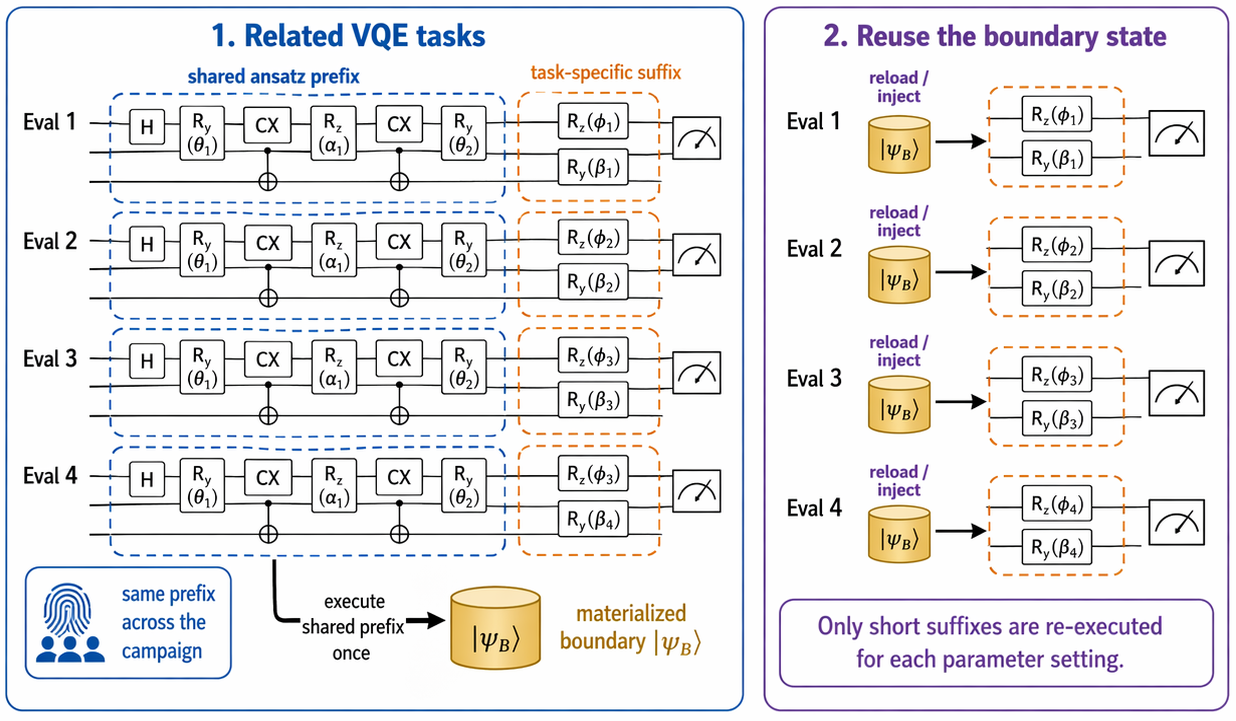}
  \caption{%
    Example workload-level reuse pattern. Four related VQE evaluations share the same ansatz prefix but differ only in short parameter-specific suffixes.
    QuWARP executes the shared prefix once, materializes the boundary state $|\psi_B\rangle$, and reloads or injects it before each suffix so only the short task-specific suffixes are re-executed.%
  }
  \label{fig:workload-example}
\end{figure}

Figure~\ref{fig:workload-example} shows the motivating case: several related VQE evaluations share the same ansatz prefix but differ only in short parameter-specific suffixes. In that setting, the useful reuse action is to execute the shared prefix once, materialize a reusable boundary state, and re-execute only the task-specific suffixes.

To address this redundancy we propose QuWARP, a bounded Qrack~\cite{strano2024qrack} execution surface that treats the workload, not the individual circuit, as the optimisation unit. It turns exactly this prefix/suffix pattern into a planning problem: related tasks are grouped, shared prefixes or segments are detected, materialize-versus-recompute cost is compared under an explicit denominator, and a boundary state is reused only when a legality guard confirms that the continuation stays on the currently supported exact surface. This legality-aware reuse matters because reusing the wrong boundary across backend or representation boundaries would be fast but incorrect. Unsupported or marginal cases are therefore not forced through reuse: the planner can refuse or abstain, and each decision is recorded in an EXPLAIN trace, meaning an inspectable report that records chosen plans, rejected alternatives, refusal reasons, provenance, and realized costs~\cite{selinger1979access,graefe1993volcano}.

Our contributions are threefold: (1) within this bounded Qrack setting, QuWARP reframes repeated-run simulation as a workload-level planning problem rather than a per-circuit routing problem; (2) it makes reuse explicit through typed exact artifacts, provenance, and a bounded legality guardrail so the evaluated handoffs are auditable rather than opaque; and (3) it shows on real quantum application workloads, spanning workflow-grounded proxies, controlled mechanism-isolation templates, and a negative control, that this bounded mechanism yields $2.95\times$--$32.84\times$ speedups over direct per-task Qrack while abstaining on the negative controls that have no shared segments. Direct per-task Qrack remains the primary comparison. Stronger but slice-specific sanity checks, including best single-context Qrack, Aer~\cite{qiskit2024}, DDSIM~\cite{zulehner2019mqt}, Qulacs~\cite{suzuki2021qulacs}, and a TQSim-style reuse reproduction~\cite{tqsim}, are reported separately so specialized stress tests do not replace the core causal comparison. The present evidence is intentionally narrower than a general orchestration claim: the validated reuse surface is Qrack's statevector mode, which stores the full amplitude vector, and its stabilizer-hybrid mode, which accelerates Clifford-dominated regions, and the workloads are chosen to mix recognizable repeated-run patterns with controlled templates rather than to claim a production-trace benchmark suite.

\section{Background and Related Work}
\label{sec:related-work}

This section places QuWARP relative to prior simulator representations, repeated-run reuse systems, and database-inspired optimisation work. That comparison matters because the paper's claim is about a different optimisation unit: campaign-level planning over related runs rather than better execution of one run at a time. Building on the Introduction, the goal here is to show why workload-level legality-aware reuse is not already provided by existing routing, caching, or simulator-selection approaches.

No single simulator representation dominates all workloads: statevector simulation, which stores the full amplitude vector, stabilizer simulation, which tracks Clifford structure efficiently, tensor-network methods, and decision-diagram methods each win on different structure~\cite{aaronson2004stabilizer, vidal2003efficient, zulehner2019mqt, vinkhuijzen2023limdd, markov2008simulating, pan2022simulation, orus2019tensor, gidney2021stim, xu2023herculeantaskclassicalsimulation, cicero2025simreview}. Prior work such as GraFeyn~\cite{grafeyn2024} reinforce that lesson by showing large gains from structure-specific sparse simulation rather than from one universal execution engine. QuWARP follows the same systems posture, but adds a workload-level planning layer above Qrack and asks when related tasks should share exact reusable state on that bounded substrate.

\begin{table}[t]
\centering
\caption{Closest-system differentiation on the evaluated Qrack surface. The rows show each system's optimisation unit and the main capability missing relative to QuWARP: workload-level typed exact reuse with legality-gated continuation and abstention.}
\label{tab:differentiation}
\footnotesize
\setlength{\tabcolsep}{4pt}
\begin{tabularx}{\columnwidth}{@{}l l X@{}}
\toprule
\textbf{System} & \textbf{Optimization unit} & \textbf{QuWARP comparison} \\
\midrule
Qrack~\cite{strano2024qrack} & circuit simulation & broad representation support; no workload-level reuse\\
Maestro~\cite{maestro2024} & circuit backend selection & routes circuits to suitable simulators; no cross-task reuse \\
Quetschlich et al.~\cite{quetschlich2023precompilation} & circuit compilation & compilation-time reuse for recurring instances; no execution-time simulation-state reuse\\
TQSim~\cite{tqsim} & multi-shot noisy circuit & reuses intermediate states across noisy shots/subcircuits; no campaign-level typed reuse planning \\
\textbf{QuWARP} & \textbf{related circuit workload} & \textbf{workload-level typed exact reuse; legality-aware, cost-based planning; abstention} \\
\bottomrule
\end{tabularx}
\end{table}

Qrack~\cite{strano2024qrack} is QuWARP's evaluated execution substrate rather than its target: Qrack adapts within a single run, whereas QuWARP optimizes across a set of related runs on the currently validated Qrack surface. Maestro~\cite{maestro2024} likewise shows that routing matters, but its decision unit is still one circuit at a time. It does not model typed reusable boundary artifacts, legality-aware continuation, or abstention across a workload.

Quetschlich et al.~\cite{quetschlich2023precompilation} is the closest motivational analogue. It reduces repeated compilation cost by reusing gate-level pre-compilation across related circuit instances with similar structure. QuWARP addresses a narrower layer in the current paper: it reuses typed simulation artifacts at execution time within a bounded Qrack surface, checks whether the continuation is legal, and can abstain when the materialization cost will not amortize.

A useful QSYS-level way to place these systems is by asking where the reuse decision lives. Decision Diagram vs. State Vector and GraFeyn explain why representation-sensitive simulator choices matter at execution time, and QuaSi shows that simulation-driven systems work can be publishable even when the core novelty is an orchestration framework rather than a new simulator kernel. Quetschlich et al. then provides the closest repeated-structure motivation, but at the gate-compilation layer. QuWARP's niche is different: on a bounded Qrack execution surface, it treats a workload of related simulation tasks as the optimisation unit, decides whether exact boundary state should be materialized and reused, and keeps unsupported or unprofitable continuations explicit rather than implicit.

TQSim is the closest prior reuse-oriented comparator because it memoizes shared prefixes across repeated trials of one circuit family~\cite{tqsim}. QuWARP differs in optimisation unit and decision surface on the evaluated Qrack substrate: it plans across related-task campaigns, searches a bounded materialize-versus-recompute decision, carries typed artifact kind/exactness/provenance metadata, and can either continue legally, refuse an illegal handoff, or abstain when no profitable legal reuse opportunity remains. EXPLAIN traces remain a secondary audit surface rather than the core distinction. Li et al.'s DM-Sim~\cite{li2020dmsim} is better viewed as a repeated-trial execution denominator than a typed workload planner; in our comparative packet it is treated as a documented host-specific exclusion.

Circuit cutting and high-performance computing (HPC) orchestration demonstrate that boundary placement and distributed execution matter~\cite{peng2020circuitcutting, tang2021cutqc, ufrecht2023circuitknitting, tejedor2025qhpc, paler2020distributed}, while System~R, Volcano, and later optimiser surveys motivate cost-based operator choice, bounded search spaces, and inspectable plans~\cite{selinger1979access, graefe1993volcano, chaudhuri1998, avnur2000}. The quantum-data-management community has additionally explored how database management systems (DBMS) can support quantum circuit simulation~\cite{Littau2025Qymera, Trummer24}. QuWARP borrows that DBMS view in a narrow systems sense for the current Qrack-only evidence surface: candidate boundaries act like alternative physical operators, materialization is chosen only when amortization is favorable, refusal makes invalid handoffs explicit, and EXPLAIN output keeps the resulting plan inspectable rather than implicit in simulator behavior.

\section{System and Implementation}
\label{sec:system-overview}

\begin{figure*}[t]
  \centering
  \includegraphics[width=\textwidth]{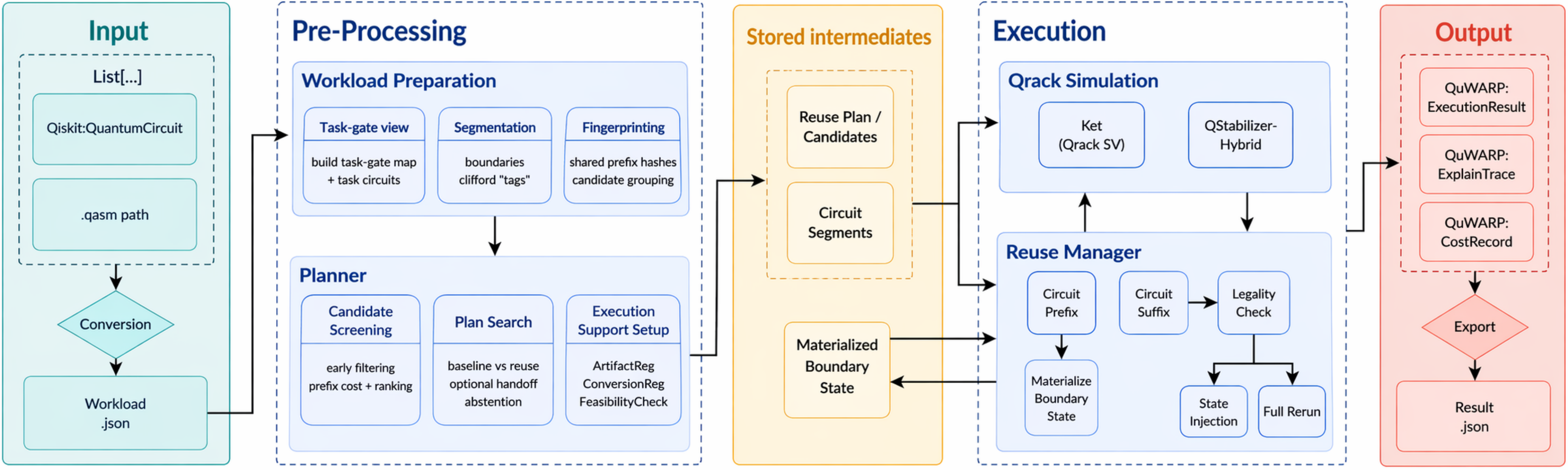}
  \caption{%
    QuWARP architecture for workload-level simulation reuse. Inputs from circuits or workload specifications are normalized into a workload representation, pre-processed into planner metadata, planned for reuse, executed on the bounded Qrack surface, and emitted together with result, EXPLAIN, and realized-cost artifacts.%
  }
  \label{fig:workload-arch}
\end{figure*}

This section presents QuWARP as a workload-planning layer over existing Qrack execution backends rather than as a new simulator kernel. Figure~\ref{fig:workload-example} gives the motivating workload-level reuse pattern, and Figure~\ref{fig:workload-arch} summarizes the full path from repeated-run inputs to planning, execution, and audit artifacts.

\subsection{Architecture Flow}

\paragraph{Input and workload preparation}
QuWARP accepts repeated-run inputs as circuit objects, QASM paths, or JSON workload descriptions and normalizes them into the workload IR used by the pipeline. The front end then prepares the planner-facing task and segment structure needed for reuse analysis without changing Qrack's simulator internals.

\paragraph{Planner and runtime support state}
The planner decomposes into candidate screening, bounded plan search, and execution-support setup. It identifies promising reuse opportunities, chooses \textsc{Reuse}, \textsc{Abstain}, or \textsc{Refuse}, and prepares the transient in-memory state needed to make the selected plan runnable.

\paragraph{Execution on the bounded Qrack surface}
When reuse is selected, the executor runs the shared circuit prefix once on the chosen backend, materializes a typed boundary state, and resumes legal suffixes from that point. Unsupported handoffs fall back to a full rerun from scratch. The evaluated backend surface remains deliberately narrow: \texttt{qrack\_sv} provides full statevector execution, while \texttt{qrack\_stabilizer} provides stabilizer-hybrid execution on Clifford-dominated regions~\cite{strano2024qrack}.

\paragraph{Outputs and audit artifacts}
Each workload emits a \texttt{QuWARP:ExecutionResult}, a \texttt{QuWARP:ExplainTrace}, and a \texttt{QuWARP:CostRecord}. Together they capture simulation outputs, planner decisions, refusal reasons, and realized costs so reuse, abstention, and fallback remain inspectable.

\label{sec:implementation}

\subsection{Implementation}

QuWARP is implemented in Python as a workload-planning layer over existing Qrack execution backends rather than as a new simulator core. Concretely, the implementation realizes Figure~\ref{fig:workload-arch}.

For evaluation Python~3.14.2 with pyqrack~1.83.1 are the core components of QuWARP. The comparative baselines are realized with qiskit~2.3.0 / qiskit-aer~0.17.2, mqt-core~3.4.1 / mqt.ddsim~2.2.0, and a Qulacs sidecar with Python~3.12.5 / qulacs~0.6.13.

Workloads arrive either as a list of \texttt{Qiskit:QuantumCircuit} or a path to a set of \textit{.qasm} files. The executor path exercised by qrack statevector or stabilizer simulation in the form of QrackSimulator in StabilizerHybrid mode. Reusable boundary states are tracked through an in-memory \texttt{ArtifactRegistry}, legality checks run before any continuation, and boundary artifacts live in RAM for the duration of a workload run. This scope point matters because the present evidence is about exact in-memory boundary reuse under planner control, not about a general on-disk storage tier.

In the current end-to-end evaluation, simulation results and intermediates are returned in JSON format this includes EXPLAIN and provenance artifacts after execution.


\section{Methods}
\label{sec:qce-methods}

This section defines the workload abstraction, legality model, and cost rule that QuWARP uses to decide when reuse is worthwhile.

\subsection{Workload IR and shared-prefix candidates}
\label{sec:workload-model}

QuWARP plans over a \emph{Workload} of related simulation \texttt{Tasks} rather than isolated runs. Figure~\ref{fig:workload-example} shows the simplest motivating case: several tasks share one large prefix and diverge only in short suffixes. Each task is cut only at coarse boundaries that matter for reuse: Clifford/non-Clifford transitions motivated by stabilizer-theory boundaries~\cite{aaronson2004stabilizer,gottesman1997stabilizer}, measurement or reset barriers, and optional user checkpoints. This keeps reuse candidates large enough to amortize state materialization while still aligning with real continuation boundaries.

\paragraph{Structural fingerprinting}
\label{sec:fingerprinting}
Each candidate segment is summarized by a deterministic hash over layout, gate names, qubit targets, and bound parameters:
\begin{equation}
\textsc{Fingerprint}(S)=\textsc{SHA256}(\textit{layout}\,\|\,g_1\,\|\,\cdots\,\|\,g_{\ell}).
\label{eq:fingerprint}
\end{equation}
Here, $S$ is one candidate segment, \textit{layout} is its qubit-layout description, and $g_1,\ldots,g_{\ell}$ are the $\ell$ ordered gate records in that segment. The equation says that QuWARP hashes the structural description of a segment into one deterministic fingerprint, so segments with the same encoded structure can be grouped quickly before expensive planning begins. Tasks whose prefix segments share the same fingerprint form a reuse candidate. Small groups, very short prefixes, and obviously cheap prefixes are rejected by configurable early filters before any detailed planning. This gives QuWARP a compact campaign-level view of where reuse is even worth considering.

\paragraph{Why coarse granularity}
Gate-level reuse would require snapshotting after almost every operation, which is more expensive than the work it saves. Coarse segmentation instead creates a small number of reuse candidates large enough to amortize. This is also the main contrast with opaque prefix-memoization schemes such as TQSim-style reuse, which cache replay points inside repeated-trial workloads rather than searching bounded workload-level materialize-versus-recompute plans over typed exact artifacts~\cite{tqsim}.

\subsection{Typed exact artifacts and legality-aware continuation}
\label{sec:artifact-model}
\label{sec:artifact-legality}
\label{sec:conversion-matrix}

A reusable boundary is stored as an \texttt{ArtifactRecord} carrying its artifact kind, exactness class, provenance hash, producing backend family, and size metadata. The provenance hash prevents stale or mismatched state from being silently reused, following the same core motivation as database provenance systems that track why and where derived artifacts came from~\cite{buneman2001provenance}. The present paper deliberately keeps this surface narrow: the only executable path classes are tableau-boundary continuation within the Qrack family, exact tableau-to-statevector materialization when dense export is supported within the qubit cap, and same-family statevector reload. Everything else becomes a structured refusal, including statevector-to-tableau handoffs, attempts to consume a statevector on a stabilizer-only backend, and any artifact whose exactness class or producing backend is unknown.

\begin{table}[t]
\centering
\caption{Compact legality summary for the evaluated Qrack surface. \textsc{STAB} denotes a stabilizer/tableau boundary state and \textsc{SV} a full statevector boundary state. Each row states whether that continuation path is supported exactly or rejected with a structured refusal; the surrounding prose gives the precise path conditions and refusal triggers.}
\label{tab:conversion-matrix}
\label{tab:refusal-reasons}
\footnotesize
\setlength{\tabcolsep}{6pt}
\begin{tabular}{@{}l l@{}}
\toprule
\textbf{Path / case} & \textbf{Outcome} \\
\midrule
STAB $\rightarrow$ STAB & \textsc{Exact} \\
STAB $\rightarrow$ SV & \textsc{Exact} \\
SV $\rightarrow$ SV & \textsc{Exact} \\
SV $\rightarrow$ STAB & \textsc{Refuse} \\
SV on stabilizer-only backend & \textsc{Refuse} \\
Non-\texttt{EXACT} artifact / unknown backend & \textsc{Refuse} \\
\bottomrule
\end{tabular}
\end{table}

\paragraph{Legality-aware handoff enforcement}
\label{sec:handoff-enforcement}
Every candidate handoff passes three gates: exactness validation, backend-family validation, and conversion-rule lookup. If any gate fails, the planner emits a structured refusal and falls back to baseline execution rather than silently coercing the state into an unsupported continuation. Together with bounded candidate screening and abstention, this keeps QuWARP from silently crossing unsupported boundaries; the main planning contribution remains the workload-level materialize-versus-recompute decision rather than the size of the legality table~\cite{tqsim}.

\paragraph{Concrete machine-checkable refusal example.}
\label{sec:legality-examples-eval}
To show that the legality boundary is real and not decorative, consider the
forbidden handoff from the failure-case gallery (EVAL-7): a four-task workload
produces a \texttt{STATEVECTOR\_BOUNDARY\_STATE} artifact on
\textsc{Qrack\_Stab\_Hybrid} and attempts to reload it on the same backend.
The feasibility checker queries the conversion matrix and finds the entry
\texttt{SV${\to}$SV} on \textsc{QSH${\to}$QSH}: \textsc{Forbidden}.
It immediately emits a structured \texttt{Refusal} with the following fields,
which appear verbatim in the EXPLAIN trace:

\begin{lstlisting}
{
  "reason": "FORBIDDEN_CONVERSION",
  "detail": "SV on stabilizer-only backend not supported.",
  "source_kind": "STATEVECTOR_BOUNDARY_STATE",
  "target_kind": "STATEVECTOR_BOUNDARY_STATE",
  "source_backend": "QRACK_STAB_HYBRID",
  "target_backend": "QRACK_STAB_HYBRID"
}
\end{lstlisting}

\noindent
There is no fallback that silently coerces the statevector into a stabilizer
continuation---the executor detects the refusal, skips materialization, and
re-executes the full circuit from scratch on the correct backend.  An
opaque prefix cache would replay the statevector blindly and produce wrong
amplitudes, so QuWARP refuses the continuation. In the evaluation, this case
is paired with companion edge cases spanning legal-but-unprofitable reuse,
accuracy-driven abstention, and random-circuit refusal.

\paragraph{Positive side: QEC-repetition legal reuse.}
For contrast, the QEC-repetition family (EVAL-3) produces a
\texttt{STATEVECTOR\_BOUNDARY\_STATE} artifact on \textsc{Qrack\_General}
and reloads it on the same backend---the \texttt{SV${\to}$SV on QG${\to}$QG:
\textsc{Exact}} entry in the conversion matrix approves the handoff
without any refusal.  The workload achieves $3.3\times$ realized speedup
with 0.51~ms planning overhead against a 2.94~ms baseline,
demonstrating that the legality layer imposes near-zero cost on
well-formed workloads while enforcing a real correctness boundary
when it counts.

\paragraph{Evidence summary.}
\label{sec:evidence-legality}
This legality surface is backed by an explicit but intentionally bounded evidence suite: 57 deterministic tests cover the forbidden matrix entries, exactness guards, runtime constraints, and refusal-trace invariants for every executed path. As shown in Section~\ref{sec:results}, especially the targeted legality ablation in Figure~\ref{fig:component-ablation}, these checks stay below $0.3{\times}$ overhead on well-formed workloads while remaining decisive on illegal continuations.

\subsection{Bounded materialize-versus-recompute planning}
\label{sec:planning-model}
\label{sec:reuse-planner}
\label{sec:cost-model}

For a candidate shared by $N$ tasks, the planner compares direct reruns against executing the prefix once, materializing the boundary, and reloading it for each continuation. Reuse is adopted only when the saved prefix work exceeds the one-time snapshot cost, reload cost, optional cross-backend handoff cost, and a safety margin. This is a deliberately compact cost-based optimisation rule in the Selinger/Volcano spirit, specialised to workload reuse rather than relational join ordering~\cite{selinger1979access,graefe1993volcano}:
\begin{equation}
(N-1)\,C_{\text{prefix}} > C_{\text{materialize}} + N\,(C_{\text{reload}} + C_{\text{handoff}}) + \delta.
\label{eq:reuse-decision}
\end{equation}
Here, $N$ is the number of tasks that share the candidate prefix, $C_{\text{prefix}}$ is the cost of executing that prefix once, $C_{\text{materialize}}$ is the one-time cost of writing the reusable boundary artifact, $C_{\text{reload}}$ is the per-task cost of reloading the artifact, $C_{\text{handoff}}$ is any extra per-task cost of a legal backend handoff, and $\delta$ is a safety margin that pushes borderline cases toward abstention. For same-backend continuation, $C_{\text{handoff}}=0$. Intuitively, the inequality says that reuse is worthwhile only when the total prefix work avoided across the remaining $N-1$ tasks exceeds all reuse overheads by a clear margin. The inequality is deliberately conservative so marginal cases fall back to direct execution rather than over-claiming reuse benefit. This is the core value of campaign-level planning: it converts many repeated runs into one explicit profitability test instead of assuming that every shared prefix should be cached.

The search itself stays bounded in three steps. First, screening keeps only the most promising candidates and proposes at most three boundary types per candidate (largest common prefix, Clifford-safe prefix, and structural divergence). Second, the planner compares a small finite set of alternatives---baseline, same-family reuse, and legal cross-backend handoff---instead of searching an open-ended plan space, again following the bounded-plan ethos of classical optimiser search spaces~\cite{selinger1979access,graefe1993volcano}. Third, the planner may abstain entirely when no candidate survives, the expected gain is too small, or the planning budget would be exceeded.

\paragraph{Abstention and complexity.}
\label{sec:abstention}
\label{sec:complexity}
Because segmentation and fingerprinting scan each task once, practical planning cost is dominated by workload construction and remains linear in the total gate count, with only a small bounded comparison term on top. This is why the \texttt{no\_share} controls can exit on the fast path with negligible overhead while reuse-positive families still realize multi-task speedups.

Reuse is attempted only when a shared boundary is structurally present, legal to continue, and predicted to amortize; otherwise the planner abstains. Section~\ref{sec:experimental-setup} next explains how that claim is tested across the six workload families.

\section{Experimental Methodology}
\label{sec:experimental-setup}

This section explains how the workloads, baselines, and measurements are constructed for the paper's empirical claims. It matters because the results should be read as controlled workload-level evidence against a fixed direct-Qrack denominator, not as ad hoc anecdotes. Building on the system and methods sections, the discussion below makes clear what is measured, on which workload families, and under which constraints.

\subsection{Hardware}

All experiments were run on a MacBook~Pro with an Apple~M3~Pro processor (up to 4.05 GHz), 12 CPU cores total, and 18\,GB unified memory, on macOS~26.4.1. All reported runs use CPU execution only, with no GPU acceleration.
Each evaluation run records its git SHA, timestamp, and provenance metadata in a manifest so every table, figure, and EXPLAIN trace can be traced back to one concrete run configuration.

\subsection{Backend surface}

The evaluated QuWARP runtime exposes two planner-selectable backends:
\begin{itemize}
  \item \texttt{qrack\_sv}---Qrack full-statevector simulation~\cite{strano2024qrack},
  \item \texttt{qrack\_stabilizer}---Qrack stabilizer-hybrid simulation~\cite{strano2024qrack}.
\end{itemize}
Aer, Qulacs, and DDSIM appear only in the separate comparative-baseline packet reported in Section~\ref{sec:comparative-baselines}; they are not planner-selected QuWARP execution backends.
Unless stated otherwise, the reuse planner uses the cost-based planning mode with a fixed calibration store.

\subsection{Workload families}

A deterministic evaluation harness regenerates the workloads, executes the configured baseline families, and exports timing tables, EXPLAIN traces, and reproducibility artifacts from one end-to-end run.

\subsection{Measurement protocol}

Each workload configuration is run with 5 repeated measurements per baseline type.
All timing values reported in Section~\ref{sec:results} are medians across repeats.
Per-task wall times are collected via Python's \texttt{time.perf\_counter\_ns} and include Qrack backend invocation~\cite{strano2024qrack}, artifact materialization/reload, and planner overhead as separate line items.

\subsection{Reproducibility}

The retained evidence bundle includes timing CSVs, EXPLAIN traces, calibration summaries, and provenance manifests sufficient to reproduce the tables and figures reported in this work.

\label{sec:evaluation}
This section completes the experimental methodology by defining the workload crosswalk, baselines, and collected metrics that support the reuse, comparison, overhead, abstention, and auditability claims.

\subsection{Workload families and justification}
Six workload families span the reuse, abstention, legality, and realism dimensions of the contribution. They are intentionally mixed rather than presented as six production traces: three are workflow-grounded proxies, two are controlled templates that isolate the reuse mechanism, and one is a negative control with no shareable structure.
Table~\ref{tab:workload-crosswalk} maps each family to the workflow it approximates, or the controlled-template role it serves, together with the reason reuse is structurally expected (or absent) and the role it plays in the evaluation story.

\begin{table*}[t]
  \centering
  \small
  \setlength{\tabcolsep}{5pt}
  \caption{%
    Workload crosswalk for the evaluation. The columns name each family, the workflow proxy or controlled-template role it represents, and the purpose it serves. The \texttt{no\_share} row is the negative control and should trigger abstention rather than reuse.%
  }
  \label{tab:workload-crosswalk}
  \begin{tabularx}{\textwidth}{@{}>{\raggedright\arraybackslash}p{3.1cm}>{\raggedright\arraybackslash}p{5.6cm}>{\raggedright\arraybackslash}X@{}}
    \toprule
    Workload Family & Workflow / template & Evaluation role \\
    \midrule
    \texttt{prefix\_heavy} &
      Controlled shared-prefix template for repeated state-preparation + readout-calibration loops~\cite{peruzzo2014vqe,qiskit2024} &
      Controlled amortization slice; isolates materialize-vs-recompute growth with $N$ \\
    \addlinespace
    \texttt{param\_sweep} &
      Controlled VQE / quantum approximate optimization algorithm (QAOA)-style ansatz-sweep template~\cite{peruzzo2014vqe,farhi2014qaoa}: shared skeleton with drifting late parameters &
      Controlled parameter-study reuse without full optimiser dynamics \\
    \addlinespace
    \texttt{vqe\_real} &
      Workflow-shaped hardware-efficient VQE energy-evaluation workload for H$_2$-style chemistry instances~\cite{peruzzo2014vqe}: repeated expectation evaluation under one ansatz family with parameter-shift updates &
      More realistic variational reuse case; $8.87\times$ at $N{=}12$ and $14.73\times$ at $N{=}24$ vs. direct Qrack \\
    \addlinespace
    \texttt{noisy\_multishot} &
      Workflow-shaped repeated noisy-sampling / shot-noise-averaging proxy~\cite{qiskit2024,tqsim} &
      High-$N$ workflow-grounded reuse; $24.81$--$32.84\times$ via single materialization \\
    \addlinespace
    \texttt{qec\_repetition} &
      Workflow-shaped QEC syndrome-cycle proxy with intermittent non-Clifford corrections~\cite{fowler2012surfacecodes} &
      Workflow-grounded legality story across stabilizer/statevector handoff \\
    \addlinespace
    \texttt{no\_share} &
      Negative control: batch of unrelated circuits (e.g., mixed quantum-volume experiments~\cite{qiskit2024}) &
      Abstention guardrail; 100\% abstention and 4~$\mu$s overhead confirm no fabricated reuse \\
    \bottomrule
  \end{tabularx}
\end{table*}

\noindent
Three of the five reuse-positive families---\texttt{vqe\_real}, \texttt{noisy\_multishot}, and \texttt{qec\_repetition}---are meant to be recognizable repeated-run workflow proxies rather than bare prefix toys.
The remaining two (\texttt{prefix\_heavy} and \texttt{param\_sweep}) are controlled templates that isolate the amortization mechanism and decision boundary without pretending to be end-to-end application traces.
The \texttt{no\_share} family is a deliberate negative control: a batch of structurally unrelated circuits ensures that QuWARP cannot accumulate wins by exploiting accidental shared structure. Its 100\% abstention outcome with bounded 4~$\mu$s overhead is as important to the whole story as the largest speedup row.

The workflow-grounded proxies were chosen to instantiate three recurring repeated-run motifs that quantum systems software actually sees: repeated expectation evaluation under a fixed ansatz family (\texttt{vqe\_real}), high-shot repeated sampling under one circuit skeleton (\texttt{noisy\_multishot}), and repeated syndrome / correction cycles with a legality-sensitive continuation boundary (\texttt{qec\_repetition}). They are therefore not presented as production traces, but as recognizable workload shapes on which shared-prefix detection, legality checks, and abstention behavior can be evaluated without conflating the planner with application-specific orchestration logic. This crosswalk matters because it ties each reported result back to one clear role in the overall evidence story.

Beyond the six primary families, the higher-cost scaling check (Section~\ref{sec:large-workloads}) extends the evaluation to absolute runtimes of 5--22 seconds per workload (GHZ-20 for 20000-qubit Greenberger--Horne--Zeilinger state preparation, random-12 with 500 gates, and quantum Fourier transform (QFT) circuits at 8/12/16 qubits). Its role is not to manufacture additional wins, but to show the realism boundary clearly: across these higher-cost workloads, the honest planner behavior is abstention when reuse still does not amortize. For the statevector-heavy cases in that set---especially random-12 and the larger QFT instances---the one-time boundary cost remains too high at the tested workload sizes.

\subsection{Baselines}
Each run compares five baseline variants so that QuWARP's contributions can be isolated:
\begin{enumerate}
  \item \emph{Direct Qrack execution} (no planner).
  \item \emph{Naive rerun} (planner infrastructure skips reuse and just reruns every task).
  \item \emph{QuWARP reuse} (full pipeline with legality checks and abstention).
  \item \emph{Legality-disabled ablation} (reuse path with legality checks turned off, evaluated on the illegal-continuation case so the refusal contribution is actually exercised).
  \item \emph{Threshold-0 ablation} (planner never abstains; we evaluate it on a near-break-even workload where the 10\% guard would otherwise keep the direct baseline).
\end{enumerate}
These comparisons isolate when reuse helps, when the planner deliberately declines a marginal plan, and when legality-aware continuation changes the outcome.

These five runs form the internal causal-isolation suite. The stronger external denominators are reported separately in Section~\ref{sec:results} because they are not uniformly applicable across every family: the Aer density-matrix path~\cite{qiskit2024} is only meaningful on \texttt{noisy\_multishot}, the current Qulacs adapter~\cite{suzuki2021qulacs} explicitly refuses measurement-bearing QEC exports, the TQSim-style reproduction~\cite{tqsim} is a reuse-vs-reuse slice rather than a plain direct baseline, and Li et al.'s DM-Sim~\cite{li2020dmsim} remains unavailable on this host. Keeping that package separate avoids conflating internal ablations with external denominators.

\subsection{Collected metrics}
We collect metrics that map directly to our core empirical claims:
\begin{itemize}
  \item Total workload wall time per family and baseline, measuring reuse speedup (Table~\ref{tab:speedup-comparison}).
  \item Planning overhead (segmenter + detector + planner + EXPLAIN export) and artifact overhead (materialization + reload), bounding the extra cost.
  \item Realized vs.\ estimated speedup for calibration assessment and auditability.
  \item Abstention and fallback counts with refusal summaries for bounded-overhead and legality evidence.
  \item Break-even thresholds $N^*$ validating the cost model's decision boundary.
\end{itemize}
All metrics are recorded per workload family.

\subsection{Ablation design and failure galleries}
The evaluation includes systematic ablations and explicit failure cases. It is structured around six questions:
\begin{enumerate}
  \item[\textbf{Q1.}] Does QuWARP exploit legal shared structure to accelerate reuse-positive workloads against the direct per-task Qrack denominator?
  \item[\textbf{Q2.}] Do stronger or more specialized comparators overturn that main result on the slices where they are semantically applicable?
  \item[\textbf{Q3.}] Are planning, materialization, and reload overheads small enough, and does the planner abstain cleanly when reuse is absent or unprofitable?
  \item[\textbf{Q4.}] Does the conservative cost model still place workloads on the correct side of the break-even boundary?
  \item[\textbf{Q5.}] On higher-cost workloads, does QuWARP preserve the same honest abstention boundary rather than manufacturing wins?
  \item[\textbf{Q6.}] Do legality checks and structured refusals prevent unsafe continuations that would otherwise appear faster but be incorrect?
\end{enumerate}
Those comparisons are summarized in Section~\ref{sec:results} in the same Q1--Q6 order, using the ablation plots, failure-case exhibit, and representative result tables to answer each question explicitly.

\section{Results}
\label{sec:results}

This section reports the empirical evidence for QuWARP's workload-level reuse claim. It matters because the paper's contribution is only useful if reuse speeds up workloads with shared structure, abstains on workloads without reusable structure, and remains inspectable at the decision boundary. Building on Section~\ref{sec:experimental-setup}, the results answer Q1--Q6 in order: reuse effectiveness, stronger baselines, overhead and abstention, break-even and calibration, the large-workload honesty boundary, and legality on the unsafe case.

\subsection{Reuse Effectiveness (Q1)}

First of all, we assess whether QuWARP exploits legal shared structure against the main direct per-task Qrack denominator. Table~\ref{tab:speedup-comparison} confirms this. Each row retains one representative slice for one workload family so that the main text shows the full mix of workflow-grounded cases, controlled templates, and the negative control without repeating every size point. The table is sorted by realized speedup from highest gain to lowest gain, with the negative control last.

The main message of Table~\ref{tab:speedup-comparison} is straightforward: every reuse-positive family lands between $3.61\times$ and $32.84\times$, while the negative-control \texttt{no\_share} family remains at $1.00\times$ with 100\% abstention. Three families are workflow-grounded repeated-run proxies, two are controlled templates, and one is the negative control; the workflow-shaped set includes a hardware-efficient H$_2$-style VQE workload rather than only a generic variational proxy. Table~\ref{tab:scale-abs-times} then follows the same row order so the reader can map each representative slice directly to its workload scale and absolute wall-clock magnitude.

\begin{table}[t]
  \centering
  \small
  \caption{Representative family-level results against direct per-task Qrack wall time. Each row is one representative workload slice from the mixed evaluation set; Speedup is realized end-to-end workload improvement, Est. is the planner's pre-execution speedup estimate, and Abstain is the abstention rate on that slice. The slice symbols indicate workload size for the family ($N$ tasks, $M$ noisy shots, or $R$ repetition rounds).}
  \label{tab:speedup-comparison}
  \setlength{\tabcolsep}{3.5pt}
  \begin{tabular}{@{}llrrr@{}}
    \toprule
    Workload Family & Representative slice & Speedup & Est. & Abstain \\
    \midrule
    noisy\_multishot & $M{=}1000$ & $\mathbf{32.84\times}$ & $5.95\times$ & 0\% \\
    vqe\_real & $N{=}24$ & $14.73\times$ & $3.40\times$ & 0\% \\
    param\_sweep & $N{=}50$ & $13.70\times$ & $2.73\times$ & 0\% \\
    prefix\_heavy & $N{=}300$ & $4.40\times$ & $3.95\times$ & 0\% \\
    qec\_repetition & $R{=}50$ & $3.61\times$ & $2.59\times$ & 0\% \\
    no\_share & $N{=}100$ & $1.00\times$ & $1.00\times$ & 100\% \\
    \bottomrule
  \end{tabular}
\end{table}

\begin{table*}[t]
  \centering
  \scriptsize
  \setlength{\tabcolsep}{4pt}
  \caption{Representative retained rows with workload scale and median absolute wall time.}
  \label{tab:scale-abs-times}
  \begin{tabularx}{\textwidth}{@{}l l r r >{\raggedright\arraybackslash}X r r >{\raggedright\arraybackslash}X@{}}
    \toprule
    Workload Family & Slice & Qubits & Gates/task & Scale descriptor & Direct (ms) & QuWARP (ms) & Role \\
    \midrule
    noisy\_multishot & $M{=}1000$ & 4 & 30 & 30-gate shared Clifford circuit repeated across 1000 shots & 108.248 & 3.296 & Workflow-grounded repeated noisy sampling family \\
    vqe\_real & $N{=}24$ & 4 & 35 & 4-qubit, 2-layer hardware-efficient VQE ansatz with 16 parameters & 3.271 & 0.222 & Workflow-grounded H$_2$-style VQE workload \\
    param\_sweep & $N{=}50$ & 4 & 15 & 10-gate shared Clifford prefix + 5 parameterized $R_z$ gates & 3.261 & 0.238 & Controlled parameterized template \\
    prefix\_heavy & $N{=}300$ & 4 & 60 & 50-gate shared Clifford prefix + 10-gate divergent T/H suffix & 61.272 & 13.911 & Controlled positive-control template \\
    qec\_repetition & $R{=}50$ & 4 & 41 & 30-gate stabilizer prefix + measure boundary + 10-gate correction suffix & 6.558 & 1.816 & Workflow-grounded QEC-style repeated syndrome / correction family \\
    no\_share & $N{=}100$ & 8 & 30 & 30-gate fully independent random circuits; zero shared-prefix expectation & 139.425 & 139.425 & Negative control with deliberate abstention target \\
    \bottomrule
  \end{tabularx}
\end{table*}

\noindent
The controlled \texttt{prefix\_heavy} family shows the expected amortization pattern: speedup increases with campaign size, and reuse becomes profitable at the profiled break-even threshold $N^* = 2$. Across the profiled \texttt{param\_sweep} and \texttt{vqe\_real} families, the empirical and estimated break-even threshold is likewise $N^* = 2$. The takeaway is that QuWARP's gains come from predictable amortization, not from one anomalous family.

\noindent
Table~\ref{tab:scale-abs-times} shows that the retained rows are not only relative wins. Because it follows the same family order as Table~\ref{tab:speedup-comparison}, the reader can compare each representative slice row by row. On these representative slices, direct workload wall time spans 3.261~ms--139.425~ms; QuWARP reduces the reuse-positive rows to 0.222--13.911~ms while leaving the negative-control \texttt{no\_share} row unchanged. Across the same retained rows, five-run variability remains small: coefficients of variation stay within 0.38\%--2.33\%, the largest interquartile range is 1.14~ms, and abstention remains fixed at 100\% on \texttt{no\_share}. The in-paper medians are therefore stable enough that the representative rows are not being driven by run-to-run noise.

\subsection{Stronger Baselines (Q2)}
\label{sec:comparative-baselines}

Next, we bserve stronger or more specialized comparators overturn the main direct-Qrack result on the slices where they actually apply. On their applicable slices, QuWARP still beats the best single-context Qrack oracle by $6.48\times$--$24.07\times$, Aer statevector by $9.91\times$--$44.16\times$, the Aer density-matrix surrogate by $31.06\times$--$33.66\times$ on \texttt{noisy\_multishot}, DDSIM by $12.76\times$--$79.24\times$ on $12/14$ workloads, Qulacs by $4.10\times$--$19.68\times$ on the measurement-free slice, and the TQSim-style/Qulacs reproduction by $1.37\times$--$5.94\times$ on the shared-prefix slice. These are deliberately secondary comparisons, not replacement denominators. The loss cases and scope limits remain visible as well: \texttt{no\_share} stays a loss, Qulacs skips measurement-bearing QEC exports, and Li et al.'s DM-Sim remains unavailable on this host.

\begin{table}[t]
  \centering
  \scriptsize
  \caption{Baseline summary on each comparator's applicable slice.}
  \label{tab:secondary-baselines}
  \begin{tabularx}{\columnwidth}{@{}>{\raggedright\arraybackslash}p{0.26\columnwidth}>{\raggedright\arraybackslash}p{0.23\columnwidth}X@{}}
    \toprule
    Comparator & Applies to & Restriction / takeaway \\
    \midrule
    Best single-context Qrack & all 14 core families & QuWARP still wins by $6.48\times$--$24.07\times$; this is a stronger internal no-reuse oracle, not workload-level reuse. \\
    Aer statevector / density matrix & exact/noiseless slices; noisy multishot only for density matrix & QuWARP stays ahead by $9.91\times$--$44.16\times$ on exact-state slices and $31.06\times$--$33.66\times$ on the noisy multishot slice; the density-matrix row is a local noisy-slice surrogate rather than a Li et al. DM-Sim reproduction. \\
    DDSIM / Qulacs & DDSIM all 14 core families; Qulacs 11 measurement-free cases & QuWARP remains ahead on their applicable slices; Qulacs skips the 3 QEC-style cases because those exports include measurement/reset structure outside the retained Qulacs slice. \\
    TQSim-style prefix reuse & 9 measurement-free shared-prefix cases & Even after giving the comparator opaque prefix memoization, QuWARP still wins by $1.37\times$--$5.94\times$; this is a reuse-vs-reuse slice, not a typed legality-aware planner. \\
    \bottomrule
  \end{tabularx}
\end{table}

\noindent
Li et al.'s DM-Sim remains an explicit host-specific exclusion rather than a plotted denominator because the cited public path is NVIDIA-GPU / cluster oriented and not meaningfully reproducible on the Apple Silicon host used for this paper.

\paragraph{Workload-level planning vs.~per-circuit steering.}
QuWARP is not a glorified per-circuit backend selector: it is a planner-based workload optimiser that processes related tasks jointly. The workload model groups related tasks into one workload-level dependency structure, and the planner produces multi-task reuse plans where the shared prefix is executed once and its boundary artifact is reloaded $N-1$ times. For example, the noisy multi-shot family at $M{=}1000$ achieves $32.84\times$ realized speedup over direct-Qrack by materializing one boundary artifact and reusing it across 1000 continuation tasks, a pattern with no analogue in per-circuit backend selection.

\paragraph{Typed artifacts vs.~opaque caches.}
Every boundary artifact carries a typed \texttt{ArtifactRecord} with an exactness class (\texttt{EXACT}/\texttt{FORBIDDEN}), provenance hash, and backend-family metadata; opaque caches carry none of this. Just as importantly, the planner does not memoize every shared prefix: it screens candidates, compares baseline execution against same-family reuse and legal handoff alternatives, and abstains when the expected gain is too small. The three-stage feasibility checker refuses illegal continuations with structured reasons. For example, the failure-case gallery shows a concrete \texttt{FORBIDDEN\_CONVERSION} refusal when a statevector artifact is sent to a stabilizer-only backend---a scenario where an opaque cache would silently produce incorrect results. An opaque prefix cache would replay the statevector into an incompatible backend silently; QuWARP's typed artifact model and feasibility layer make that impossible by construction.

\subsection{Overhead and Abstention (Q3)}

To evaluate QuWARP's benefits further it is crucial to check whether planner and artifact bookkeeping remain small relative to the avoided repeated prefix work, and whether the system exits cleanly when reuse is not warranted. Table~\ref{tab:planning-overhead} shows that both conditions hold: materialization stays sub-millisecond, reload remains about 2~$\mu$s per task on reuse-positive families, and the negative-control \texttt{no\_share} family exits through the fast-path abstention route in 4~$\mu$s. The main message of the table is that QuWARP's extra planning logic remains small compared with the saved repeated prefix work.

\begin{table}[t]
  \centering
  \scriptsize
  \setlength{\tabcolsep}{3pt}
  \caption{Planning and artifact-overhead by workload family. The negative-control \texttt{no\_share} row is the key sanity check: QuWARP exits through abstention with near-zero extra cost.}
  \label{tab:planning-overhead}
  \begin{tabularx}{\columnwidth}{@{}X l r r r@{}}
    \toprule
    Workload & Slice & Planning time (ms) & Boundary Mat. (ms) & Reload ($\mu \text{s}$) \\
    \midrule
    Noisy multishot & $M{=}1000$ & 19.68 & 0.011 & 2 \\
    VQE real & $N{=}24$ & 1.37 & 0.007 & 2 \\
    Prefix-heavy & $N{=}300$ & 9.76 & 0.011 & 2 \\
    QEC repetition & $R{=}50$ & 1.02 & 0.007 & 2 \\
    No share & $N{=}100$ & 0.004 & 0.0 & 0 \\
    \bottomrule
  \end{tabularx}
\end{table}

\subsection{Break-Even and Calibration (Q4)}

\begin{figure}[t]
  \centering
  \includegraphics[width=0.95\columnwidth]{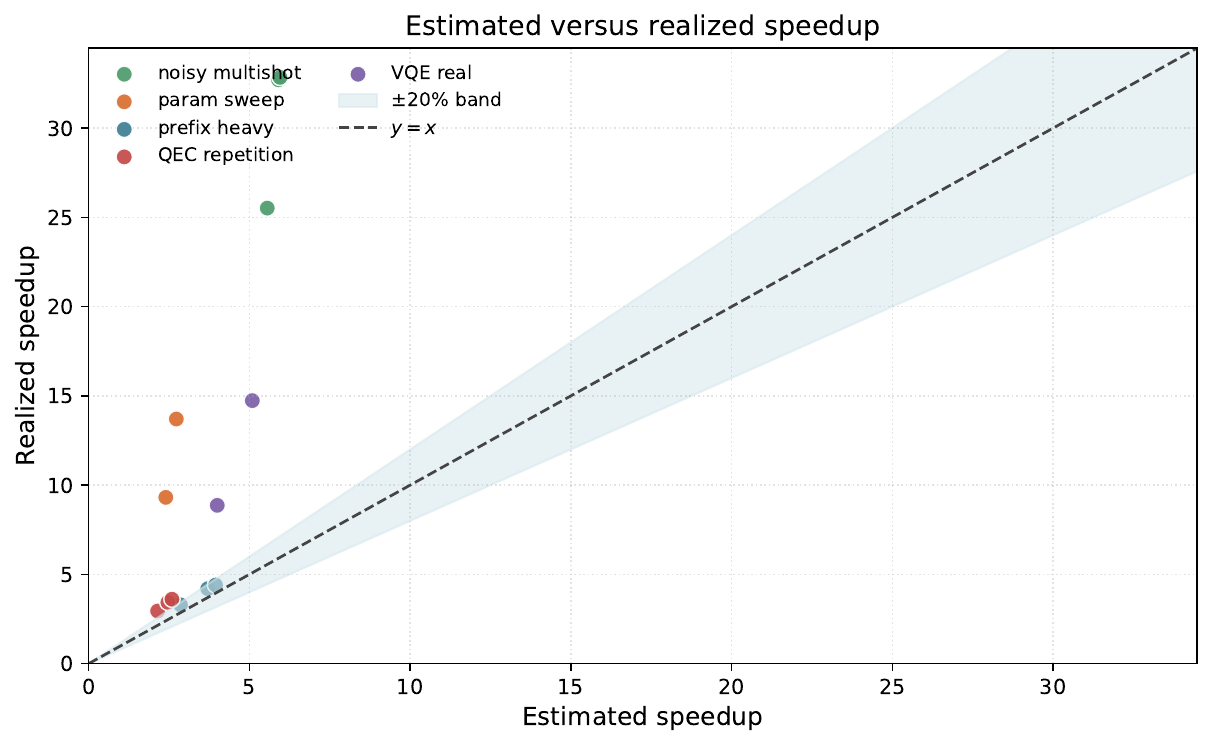}
  \caption{Estimated-versus-realized workload speedup for all evaluated workloads. Each point is one workload; the x-axis is estimated speedup and the y-axis is realized speedup against direct per-task Qrack. The diagonal denotes perfect calibration and the shaded band marks close agreement. All points lie above the diagonal, so the current gate-count proxy underestimates gains while still placing workloads on the correct side of the reuse boundary.}
  \label{fig:est-vs-realized}
\end{figure}

The proposed conservative cost model places workloads on the correct side of the reuse boundary. Across the profiled \texttt{prefix\_heavy}, \texttt{param\_sweep}, and \texttt{vqe\_real} families, both the empirical and estimated break-even threshold is $N^* = 2$, so reuse becomes profitable as soon as one additional continuation can amortize the one-time boundary cost. Figure~\ref{fig:est-vs-realized} shows the same boundary story at workload scale: estimated speedups of $2.59\times$--$5.95\times$ correspond to realized speedups of $3.61\times$--$32.84\times$, so the current model is systematically conservative without placing the retained workloads on the wrong side of the decision boundary.

What matters here is boundary correctness rather than exact magnitude prediction. The retained \texttt{vqe\_real} and \texttt{qec\_repetition} rows show that point on workflow-grounded families rather than only on near-identical repeats: the VQE workload estimates $3.40\times$ and realizes $14.73\times$, while the QEC repetition workload estimates $2.59\times$ and realizes $3.61\times$. The EXPLAIN bundle preserves the selected boundary, rejected alternatives, refusal reasons, and estimated-versus-realized costs, so this conservative calibration behavior remains auditable rather than implicit.

\subsection{Large-Workload Honesty Boundary (Q5)}
\label{sec:large-workloads}

QuWARP succesfully preserves the same honest abstention boundary once absolute runtimes reach seconds rather than milliseconds. GHZ-20 (20,000 qubits), random-12 with 500 gates, and QFT scaling at 8/12/16 qubits all preserve the same outcome: QuWARP abstains when reuse does not yet amortize at the tested workload sizes. For the statevector-heavy members of this set---notably random-12 and the larger QFT instances---the one-time boundary cost remains too high; GHZ-20 serves instead as a higher-cost Clifford-side stress case. These second-scale runs are included precisely to expose the realism boundary, not to manufacture extra wins. Their role is to show where the present bounded reuse surface stops helping, which is just as important as the positive rows in Table~\ref{tab:speedup-comparison}.

\subsection{Legality and Unsafe Ablation (Q6)}

\begin{figure}[t]
  \centering
  \includegraphics[width=\columnwidth]{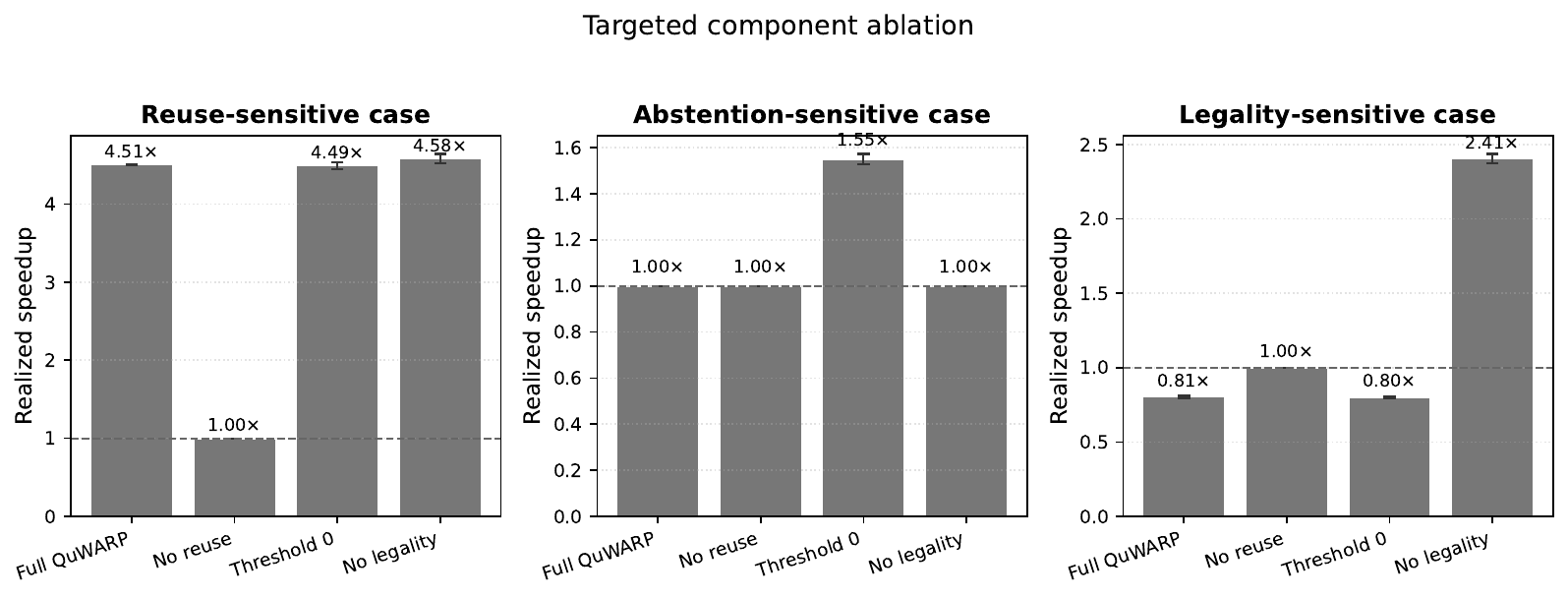}
  \caption{Targeted planner-component ablation on the cases where each mechanism matters. Removing reuse collapses the reuse-positive workload to the direct baseline; removing the abstention gate admits a marginal near-break-even plan; removing legality on the illegal-continuation case permits an unsafe unsupported handoff instead of QuWARP's refusal and fallback. Error bars show five reruns.}
  \label{fig:component-ablation}
\end{figure}

Finally, legality-aware refusal matters because the issue is correctness, not only speed. Figure~\ref{fig:component-ablation} reports a targeted component ablation on the cases where each planner mechanism is actually supposed to matter. On the reuse-positive prefix-heavy workload, removing reuse collapses speedup from $4.51\times$ to the $1.00\times$ direct baseline, while legality and abstention stay essentially flat on an already-legal path. On the near-break-even prefix-heavy case, the 10\% abstention gate deliberately keeps the direct baseline at $1.00\times$, whereas the threshold-0 variant admits a small $1.55\times$ win; this is a conservative policy choice, not a missing component effect.

The illegal-continuation case is different. There, the unsafe path is not merely faster bookkeeping; it permits an illegal continuation by replaying a boundary artifact in an unsupported execution context, such as handing a statevector artifact to a stabilizer-only continuation/backend path. That can produce incorrect or non-exact results. Full QuWARP therefore checks the handoff, emits a structured refusal, and falls back to rerunning from scratch on a supported path. On this intentionally bad case, full QuWARP can be slower than the plain direct baseline because it pays the extra cost of checking, refusing, and then rerunning, whereas the direct baseline never attempts the illegal reuse path at all and simply executes from scratch once. The legality-disabled bar is therefore not a better operating point; it is an unsafe one. If a user knew in advance that a workload contained no legal or profitable reuse opportunities, plain baseline execution would indeed be preferable. QuWARP's contribution is that it detects and refuses unsafe reuse generically rather than silently producing the wrong answer.

\section{Discussion}
\label{sec:discussion}

This section interprets where QuWARP helps and where the current evidence stops.

\paragraph{When workload-level reuse matters}
Table~\ref{tab:speedup-comparison} and the break-even evidence in Section~\ref{sec:results} show that reuse wins only when three conditions align: a substantial shared prefix, enough tasks to amortize materialization, and a legal continuation. Those conditions hold on the hardware-efficient VQE, noisy-multishot, QEC, prefix-heavy, and parameter-sweep families, and they fail by design on \texttt{no\_share} and the large-workload abstention cases. That mixed evaluation is important: the positive rows show that workload-level reuse helps beyond toy repeats, while the explicit loss cases show that QuWARP does not manufacture wins when shared structure is absent or too expensive to reuse.

\paragraph{Positioning against TQSim}
TQSim is the closest prior computational-reuse system and therefore the right opponent for QuWARP~\cite{tqsim}. Its published target is repeated-shot statevector simulation under noise: one circuit is partitioned, intermediate states are reused across the resulting shot tree, and the main applications are noisy Monte Carlo-style studies such as the paper's QAOA parameter-grid case study. That is a meaningful reuse contribution, and our comparative packet keeps a TQSim-style reproduction explicit for exactly that reason. QuWARP, however, answers a different systems question. Its optimisation unit is a workload of related tasks rather than the shot tree of one noisy circuit. That makes QuWARP the more natural fit for repeated exact VQE energy evaluations, parameter sweeps, QEC-style syndrome/correction workflows, and mixed workload batches that contain both reuse-positive and abstention-worthy cases. The distinction also shows up empirically: even on the narrow reuse-vs-reuse slice where the TQSim-style reproduction applies, QuWARP still leads by $1.37\times$--$5.94\times$ because the planner is solving a campaign-level decision problem rather than only replaying cached prefixes.

\paragraph{Why typed exact artifacts still matter}
The cleanest way to place the two systems is therefore not simply ``TQSim for noisy, QuWARP for exact.'' The TQSim paper explicitly notes that its approach can be combined with techniques that improve single-shot simulation time~\cite{tqsim}. Even so, the published TQSim design and evaluation remain centered on multi-shot noisy statevector simulation, so the core distinction would still be the safety contract and optimisation unit rather than a binary noisy-versus-exact split. TQSim accepts an explicit speed-versus-fidelity tradeoff through its partitioning and shot-allocation choices~\cite{tqsim}. QuWARP instead stays on a bounded exact surface: reusable states are typed exact artifacts with provenance, legality is checked before continuation, and the planner may abstain or refuse rather than force reuse. When a continuation is illegal, the feasibility checker refuses it with a structured reason and the provenance hash~\cite{buneman2001provenance} prevents stale reuse even within one backend family. That boundary is intentionally modest, but it is real: the ablation packet shows the checks stay cheap ($\Delta_L < 0.3\times$) on supported workloads, while the failure-case gallery shows why removing them would silently permit incorrect continuation. In that sense the systems are better viewed as complementary layers than as interchangeable caches: TQSim optimizes reuse within a noisy repeated-shot campaign, whereas QuWARP decides when exact reuse is legal, profitable, and auditable across a workload.

\paragraph{Conservatism as a feature}

Figure~\ref{fig:est-vs-realized} shows an intentional conservative reuse policy. Although estimated speedups understate realized gains, the planner's purpose is not exact performance prediction but robust decision-making: reuse is selected only when an exact boundary is legal and the expected benefit clearly exceeds materialization and reload costs. This conservative bias helps identify the observed break-even boundary while preserving clean abstention on workloads where reuse is absent, unsupported, or not predicted to yield significant speedups.

\paragraph{Limitations}
The current evidence is limited to Qrack's statevector and stabilizer backends. The cost model is intentionally conservative and may leave some marginal reuse opportunities unused. These constraints keep the conclusions narrow: QuWARP demonstrates measurable workload-level reuse and clean abstention on exact simulation backends, but broader cross-simulator evidence using tensor networks (TN) or decision diagrams (DD) remains future work. The reason is not that TN or DD simulation is unsupported in principle, but that exact conversion into a compact TN/DD boundary states is structure-dependent and can be costly enough to eliminate the reuse benefit. Supporting these backends would therefore require backend-specific conversion rules and cost models rather than an unchecked handoff~\cite{markov2008simulating, TNTreeWidth2018,willeDD2021,vinkhuijzen2023limdd}.

\section{Conclusion}
\label{sec:conclusion}

Across six workload families on the evaluated Qrack statevector / stabilizer-hybrid backends, QuWARP delivers $2.95\times$--$32.84\times$ speedups over direct per-task Qrack on reuse-positive workloads and abstains on negative controls. The main result is that repeated-run simulation benefits from workload-level reuse planning when shared structure is present and the continuation is legal.

The claim is intentionally narrow: QuWARP validates typed exact artifacts, legality-aware continuation, and abstention on Qrack backends rather than broad simulator interoperability.

\bibliographystyle{IEEEtran}
\bibliography{refs}

@article{markov2008simulating,
  title        = {Simulating quantum computation by contracting tensor networks},
  author       = {Markov, Igor L. and Shi, Yaoyun},
  journal      = {SIAM Journal on Computing},
  volume       = {38},
  number       = {3},
  pages        = {963--981},
  year         = {2008},
  doi          = {10.1137/050644756}
}

@misc{qiskit2024,
  author       = {{Qiskit Contributors}},
  title        = {Qiskit: An Open-Source Framework for Quantum Computing},
  howpublished = {\url{https://qiskit.org}},
  year         = {2024},
  note         = {Accessed: 2026-02-12}
}

@misc{strano2024qrack,
  tauthor = { Strano, Daniel and Bollay, Benn and Blaauw, Aryan and Shammah, Nathan and Zeng, William J. and Mari, Andrea },
  booktitle = { 2023 IEEE International Conference on Quantum Computing and Engineering (QCE) },
  title = {{ Exact and approximate simulation of large quantum circuits on a single GPU }},
  year = {2023},
  volume = {},
  ISSN = {},
  pages = {949-958},
  doi = {10.1109/QCE57702.2023.00109},
  url = {https://doi.ieeecomputersociety.org/10.1109/QCE57702.2023.00109},
  publisher = {IEEE Computer Society},
  address = {Los Alamitos, CA, USA},
  month =sep
}

@article{suzuki2021qulacs,
  title        = {Qulacs: a fast and versatile quantum circuit simulator for research purpose},
  author       = {Suzuki, Yasunari and Yoshioka, Shumpei and Ikeda, Takahiko and Raymond, Rudy and Tanaka, Naoki and Onodera, Takuya and Yamamoto, Hiroshi and Koide, Takashi and Morino, Akira and Sawada, Keisuke and Nakanishi, Hiroki and Mitarai, Kosuke and Fujii, Keisuke},
  journal      = {Quantum},
  volume       = {5},
  pages        = {559},
  year         = {2021},
  doi          = {10.22331/q-2021-10-06-559}
}

@article{aaronson2004stabilizer,
  title        = {Improved simulation of stabilizer circuits},
  author       = {Aaronson, Scott and Gottesman, Daniel},
  journal      = {Physical Review A},
  volume       = {70},
  number       = {5},
  pages        = {052328},
  year         = {2004},
  doi          = {10.1103/PhysRevA.70.052328}
}

@phdthesis{gottesman1997stabilizer,
  title        = {Stabilizer codes and quantum error correction},
  author       = {Gottesman, Daniel},
  school       = {California Institute of Technology},
  year         = {1997},
  note         = {arXiv:quant-ph/9705052}
}

@article{vidal2003efficient,
  title        = {Efficient classical simulation of slightly entangled quantum computations},
  author       = {Vidal, Guifr{\'e}},
  journal      = {Physical Review Letters},
  volume       = {91},
  number       = {14},
  pages        = {147902},
  year         = {2003}
}

@article{zulehner2019mqt,
  title        = {Advanced simulation of quantum computations},
  author       = {Zulehner, Alwin and Wille, Robert},
  journal      = {IEEE Transactions on Computer-Aided Design of Integrated Circuits and Systems},
  volume       = {38},
  number       = {5},
  pages        = {848--859},
  year         = {2019},
  doi          = {10.1109/TCAD.2018.2834427}
}

@article{maestro2024,
  title        = {Maestro: Intelligent Execution for Quantum Circuit Simulation},
  author       = {Bertomeu, Oriol and Ghayas, Hamzah and Roman, Adrian and DiAdamo, Stephen},
  journal      = {arXiv preprint arXiv:2512.04216},
  year         = {2025}
}

@article{peng2020circuitcutting,
  title        = {Simulating large quantum circuits on a small quantum computer},
  author       = {Peng, Tianyi and Harrow, Aram W. and Ozols, Maris and Wu, Xiaodi},
  journal      = {Physical Review Letters},
  volume       = {125},
  number       = {15},
  pages        = {150504},
  year         = {2020},
  doi          = {10.1103/PhysRevLett.125.150504}
}

@article{tang2021cutqc,
  title        = {{CutQC}: Using Small Quantum Computers for Large Quantum Circuit Evaluations},
  author       = {Tang, Wei and Tomesh, Teague and Suchara, Martin and Larson, Jeffrey and Martonosi, Margaret},
  journal      = {Proceedings of ASPLOS},
  year         = {2021}
}

@article{ufrecht2023circuitknitting,
  doi = {10.22331/q-2023-10-23-1147},
  url = {https://doi.org/10.22331/q-2023-10-23-1147},
  title = {Cutting multi-control quantum gates with {ZX} calculus},
  author = {Ufrecht, Christian and Periyasamy, Maniraman and Rietsch, Sebastian and Scherer, Daniel D. and Plinge, Axel and Mutschler, Christopher},
  journal = {{Quantum}},
  issn = {2521-327X},
  publisher = {{Verein zur F{\"{o}}rderung des Open Access Publizierens in den Quantenwissenschaften}},
  volume = {7},
  pages = {1147},
  month = oct,
  year = {2023}
}

@inproceedings{graefe1993volcano,
  title        = {The {Volcano} optimizer generator: Extensibility and efficient search},
  author       = {Graefe, Goetz and McKenna, William J.},
  booktitle    = {Proceedings of the 9th International Conference on Data Engineering (ICDE)},
  pages        = {209--218},
  year         = {1993},
  organization = {IEEE}
}

@inproceedings{selinger1979access,
  title        = {Access path selection in a relational database management system},
  author       = {Selinger, P. Griffiths and Astrahan, Morton M. and Chamberlin, Donald D. and Lorie, Raymond A. and Price, Thomas G.},
  booktitle    = {Proceedings of the 1979 ACM SIGMOD International Conference on Management of Data},
  pages        = {23--34},
  year         = {1979}
}

@inproceedings{buneman2001provenance,
  title        = {Why and Where: A Characterization of Data Provenance},
  author       = {Buneman, Peter and Khanna, Sanjeev and Tan, Wang-Chiew},
  booktitle    = {Proceedings of the 8th International Conference on Database Theory (ICDT)},
  series       = {Lecture Notes in Computer Science},
  volume       = {1973},
  pages        = {316--330},
  year         = {2001},
  doi          = {10.1007/3-540-44503-X_20}
}

@article{pan2022simulation,
  title        = {Simulation of quantum circuits using the big-batch tensor network method},
  author       = {Pan, Feng and Zhang, Pan},
  journal      = {Physical Review Letters},
  volume       = {128},
  number       = {3},
  pages        = {030501},
  year         = {2022}
}

@article{peruzzo2014vqe,
  title={A variational eigenvalue solver on a photonic quantum processor},
  author={Peruzzo, Alberto and McClean, Jarrod and Shadbolt, Peter and Yung, Man-Hong and Zhou, Xiao-Qi and Love, Peter J and Aspuru-Guzik, Al{\'a}n and O'Brien, Jeremy L},
  journal={Nature Communications},
  volume={5},
  pages={4213},
  year={2014},
  publisher={Nature Publishing Group}
}

@article{farhi2014qaoa,
  title={A quantum approximate optimization algorithm},
  author={Farhi, Edward and Goldstone, Jeffrey and Gutmann, Sam},
  journal={arXiv preprint arXiv:1411.4028},
  year={2014}
}

@article{fowler2012surfacecodes,
  title        = {Surface codes: Towards practical large-scale quantum computation},
  author       = {Fowler, Austin G. and Mariantoni, Matteo and Martinis, John M. and Cleland, Andrew N.},
  journal      = {Physical Review A},
  volume       = {86},
  number       = {3},
  pages        = {032324},
  year         = {2012},
  doi          = {10.1103/PhysRevA.86.032324}
}

@inproceedings{li2020dmsim,
  title        = {Density Matrix Quantum Circuit Simulation via the {BSP} Machine on Modern {GPU} Clusters},
  author       = {Li, Ang and Subasi, Omer and Yang, Xiu and Krishnamoorthy, Sriram},
  booktitle    = {Proceedings of the International Conference for High Performance Computing, Networking, Storage and Analysis (SC)},
  year         = {2020},
  doi          = {10.5555/3433701.3433718}
}

@inproceedings{tqsim,
  title        = {Accelerating Simulation of Quantum Circuits under Noise via Computational Reuse},
  author       = {Wang, Meng and Tannu, Swamit and Nair, Prashant J.},
  booktitle    = {Proceedings of the 52nd Annual International Symposium on Computer Architecture (ISCA)},
  year         = {2025},
  pages        = {1539--1553},
  doi          = {10.1145/3695053.3730992}
}

@article{cicero2025simreview,
  title        = {Simulation of Quantum Computers: Review and Acceleration Opportunities},
  author       = {Cicero, Alessio and Maleki, Mohammad Ali and Azhar, Muhammad Waqar and Ul-Abdin, Zain and Svensson, B. Johan},
  journal      = {ACM Computing Surveys},
  year         = {2025},
  doi          = {10.1145/3762672}
}

@inproceedings{tejedor2025qhpc,
  title        = {Orchestrating Quantum-{HPC} Workflows with Distributed Quantum Circuit Cutting},
  author       = {Tejedor, Mar and Casas, Berta and Conejero, Javier and others},
  booktitle    = {Proceedings of the 2025 International Workshop on Quantum Classical Cooperative (QCC)},
  year         = {2025},
  doi          = {10.1145/3731599.3767547}
}

@article{vinkhuijzen2023limdd,
  title={LIMDD: A Decision Diagram for Simulation of Quantum Computing including Stabilizer States},
  author={Vinkhuijzen, Lieuwe and Coopmans, Tim and Elkouss, David and Dunjko, Vedran and Laarman, Alfons},
  journal={Quantum},
  volume={7},
  pages={1108},
  year={2023},
  publisher={Verein zur F{\"o}rderung des Open Access Publizierens in den Quantenwissenschaften}
}

@article{orus2019tensor,
  title={Tensor Networks for Complex Quantum Systems},
  author={Or{\'u}s, Rom{\'a}n},
  journal={Nature Reviews Physics},
  volume={1},
  number={9},
  pages={538--550},
  year={2019},
  publisher={Nature Publishing Group UK London}
}

@misc{xu2023herculeantaskclassicalsimulation,
  title={A Herculean Task: Classical Simulation of Quantum Computers},
  author={Xiaosi Xu and Simon Benjamin and Jinzhao Sun and Xiao Yuan and Pan Zhang},
  year={2023},
  url={https://arxiv.org/abs/2302.08880}
}

@article{gidney2021stim,
  title={Stim: a fast stabilizer circuit simulator},
  author={Gidney, Craig},
  journal={Quantum},
  volume={5},
  pages={497},
  year={2021},
  publisher={Verein zur F{\"o}rderung des Open Access Publizierens in den Quantenwissenschaften}
}

@inproceedings{Littau2025Qymera,
  author = {Littau, Tim and Hai, Rihan}, 
  title = {Qymera: Simulating Quantum Circuits using RDBMS}, 
  year = {2025}, 
  isbn = {9798400715648}, 
  publisher = {Association for Computing Machinery}, 
  address = {New York, NY, USA}, 
  url = {https://doi.org/10.1145/3722212.3725126}, 
  doi = {10.1145/3722212.3725126},
  booktitle = {Companion of the 2025 International Conference on Management of Data}, pages = {179-182}, 
  numpages = {4}, 
  location = {Berlin, Germany}, 
  series = {SIGMOD/PODS '25}
}

@inproceedings{Trummer24,
author = {Trummer, Immanuel},
title = {Towards Out-of-Core Simulators for Quantum Computing},
year = {2024},
booktitle = {Proceedings of the 1st Workshop on Quantum Computing and Quantum-Inspired Technology for Data-Intensive Systems and Applications},
series = {Q-Data '24}
}

@inproceedings{chaudhuri1998, 
author = {Chaudhuri, Surajit}, 
title = {An overview of query optimization in relational systems}, 
year = {1998}, 
isbn = {0897919963}, 
publisher = {Association for Computing Machinery}, 
address = {New York, NY, USA}, 
url = {https://doi.org/10.1145/275487.275492}, 
doi = {10.1145/275487.275492}, 
booktitle = {Proceedings of the Seventeenth ACM SIGACT-SIGMOD-SIGART Symposium on Principles of Database Systems}, 
pages = {34–43}, 
numpages = {10}, 
location = {Seattle, Washington, USA}, 
series = {PODS '98} 
}

@article{avnur2000, 
author = {Avnur, Ron and Hellerstein, Joseph M.}, 
title = {Eddies: continuously adaptive query processing}, 
year = {2000}, 
issue_date = {June 2000}, 
publisher = {Association for Computing Machinery}, 
address = {New York, NY, USA}, 
volume = {29}, 
number = {2}, 
issn = {0163-5808}, 
url = {https://doi.org/10.1145/335191.335420}, 
doi = {10.1145/335191.335420}, 
journal = {SIGMOD Rec.}, 
month = may, 
pages = {261–272}, 
numpages = {12} 
}

@article{paler2020distributed,
  title={Distributed quantum circuit simulation},
  author={Paler, Alexandru and Fowler, Austin G},
  journal={Quantum Science and Technology},
  volume={5},
  number={3},
  pages={034003},
  year={2020},
  publisher={IOP Publishing}
}

@inproceedings{grafeyn2024,
  title        = {GraFeyn: Efficient Parallel Sparse Simulation of Quantum Circuits},
  author       = {Westrick, Sam and others},
  booktitle    = {2024 IEEE International Conference on Quantum Computing and Engineering (QCE)},
  pages        = {1132--1142},
  year         = {2024},
  doi          = {10.1109/QCE60285.2024.00132}
}

@inproceedings{quetschlich2023precompilation,
  title        = {Reducing the Compilation Time of Quantum Circuits Using Pre-Compilation on the Gate Level},
  author       = {Quetschlich, Nils and Burgholzer, Lukas and Wille, Robert},
  booktitle    = {2023 IEEE International Conference on Quantum Computing and Engineering (QCE)},
  pages        = {757--767},
  year         = {2023},
  doi          = {10.1109/QCE.2023.10313914}
}

@article{TNTreeWidth2018,
    doi = {10.1371/journal.pone.0207827},
    author = {Dumitrescu, Eugene F. AND Fisher, Allison L. AND Goodrich, Timothy D. AND Humble, Travis S. AND Sullivan, Blair D. AND Wright, Andrew L.},
    journal = {PLOS ONE},
    publisher = {Public Library of Science},
    title = {Benchmarking treewidth as a practical component of tensor network simulations},
    year = {2018},
    month = {12},
    volume = {13},
    url = {https://doi.org/10.1371/journal.pone.0207827},
    pages = {1-19},
    number = {12},

}

@INPROCEEDINGS{willeDD2021,
  author={Wille, Robert and Burgholzer, Lukas and Artner, Michael},
  booktitle={2021 Design, Automation \& Test in Europe Conference \& Exhibition (DATE)}, 
  title={Visualizing Decision Diagrams for Quantum Computing (Special Session Summary)}, 
  year={2021},
  volume={},
  number={},
  pages={768-773},
  doi={10.23919/DATE51398.2021.9474236}
}

\end{document}